\documentclass[aps,prb,twocolumn,superscriptaddress,floatfix,longbibliography]{revtex4-1}
\usepackage{amssymb}
\usepackage{graphicx}
\usepackage{dcolumn}
\usepackage{bm}
\usepackage{amsmath}

\usepackage[colorlinks,linkcolor=magenta,citecolor=blue,urlcolor=blue]{hyperref}
\renewcommand{\Re}{\operatorname{Re}}
\begin{document}

%\title{Mobility edge in one-dimensional non-Hermitian incommensurate lattices}
%\title{Non-Hermitian mobility edge in quasiperiodic lattices with parity-time symmetry}
\title{Non-Hermitian mobility edges in one-dimensional quasicrystals with parity-time symmetry}
%%%%%%%%authors%%%%%%%%%%%%%%%
\author{Yanxia Liu}\thanks{These authors contributed equally to this work.}
\affiliation{Beijing National Laboratory for Condensed Matter Physics, Institute of Physics, Chinese Academy of Sciences, Beijing 100190, China}
\author{Xiang-Ping Jiang}\thanks{These authors contributed equally to this work.}
\affiliation{Beijing National Laboratory for Condensed Matter Physics, Institute of Physics, Chinese Academy of Sciences, Beijing 100190, China}
\affiliation{School of Physical Sciences, University of Chinese Academy of Sciences, Beijing, 100049, China}
\author{Junpeng Cao}
\affiliation{Beijing National Laboratory for Condensed Matter Physics, Institute of Physics, Chinese Academy of Sciences, Beijing 100190, China}
\affiliation{School of Physical Sciences, University of Chinese Academy of Sciences, Beijing, 100049, China}
\affiliation{Songshan lake Materials Laboratory, Dongguan, Guangdong 523808, China}
\author{Shu Chen}
\thanks{Corresponding author: schen@iphy.ac.cn}
\affiliation{Beijing National Laboratory for Condensed Matter Physics, Institute of Physics, Chinese Academy of Sciences, Beijing 100190, China}
\affiliation{School of Physical Sciences, University of Chinese Academy of Sciences, Beijing, 100049, China}
\affiliation{Yangtze River Delta Physics Research Center, Liyang, Jiangsu 213300, China}
%%%%%%%%authors%%%%%%%%%%%%%%%

\date{\today}

\begin{abstract}
We investigate localization-delocalization transition in one-dimensional non-Hermitian quasiperiodic lattices with exponential short-range hopping, which possess parity-time ($\mathcal{PT}$) symmetry. The localization transition induced by the non-Hermitian quasiperiodic potential is found to occur at the $\mathcal{PT}$-symmetry-breaking point. Our results also demonstrate the existence of energy dependent mobility edges, which separate the extended states from localized states  and are only associated with the real part of eigen-energies. The level statistics and Loschmidt echo dynamics are also studied.
\end{abstract}

\maketitle

\section{Introduction}
Ever since the seminal work of Anderson \cite{anderson1958absence}, Anderson localization has become a fundamental paradigm for the study of localization induced by random disorder in condensed matter physics. While all eigenstates are localized in the presence of infinitesimal disorder strengths in one- and two-dimensional noninteracting systems, localized and extended states can coexist at different energies in three
dimensions with a single-particle mobility edge (SPME)
\cite{abrahams1979scaling,lee1985disordered,evers2008anderson}, i.e., a critical energy separating localized and delocalized energy
eigenstates.
%Particularly, direct observation of this phenomena in ultracold atomic systems with controlled artificial disorder has been realized
%in recent experiments \cite{roati2008anderson,billy2008direct,jendrzejewski2012three,mcgehee2013three,kondov2011three}.
As an intermediate case between the disordered and periodic systems, quasicrystals display very different behaviors and may support a localization-delocalization transition
even in one dimension.  A well known example is given by a one-dimensional (1D) quasiperiodic
system described by the Aubry-Andr\'{e} (AA)
model \cite{aubry1980analyticity,harper1955single}, which undergoes a localization transition
when the strength of the quasiperiodical potential exceeds a critical point determined by the self-duality condition. The AA model has been experimentally realized in bichromatic optical lattices \cite{roati2008anderson}. By introducing short-range or long-range hopping processes, some modified AA
models may support energy-dependent mobility edges \cite{biddle2011localization,biddle2010predicted,ganeshan2015nearest,li2016quantum,li2017mobility,li2018mobility,DengX}, which were found to appear in other quasiperiodic models \cite{sarma1988mobility,thouless1988localization,sarma1990localization,Rossignolo,GuoAM,Boers,Gadway}.
Experimental observation of mobility edge and many body localization in 1D quasiperiodic optical
lattices was also reported in recent works \cite{luschen2018single,kohlert2019observation}.

Recently, there has been growing interest in non-Hermitian Hamiltonians from theory to experiment \cite{Ozawa,Bender,Guo,Gong,LiuCH1,LiuCH2,Sato,Zhou}. In general, the non-Hermiticity is
achieved by introducing nonreciprocal hopping processes or gain and loss terms, which may induce exotic phenomena without Hermitian
counterparts, such as parity-time ($ \mathcal{PT} $) phase transitions \cite{Bender,ZhuBG,Yuce2015,Esaki}, exceptional points and non-Hermitian skin effect \cite{Yao,Alvarez,Xiong,Kunst,Yao1,Yin,Lieu,TELee,ShenH,Leykam}. Interplay of non-Hermiticity and disorder was
studied in terms of the Hatano-Nelson type models \cite{hatano1996localization,hatano1998non,feinberg1999non,kolesnikov2000localization,Nelson1998}, in which
the nonreciprocal hopping introduced in the 1D Anderson model leads to a finite localization-delocalization transition.
%and non-Hermitian Anderson models with complex on-site disorder potentials \cite{tzortzakakis2019non,Goldsheid,Molinari}.
Effects of non-Hermiticity on quasiperiodic lattices have
been studied in different contexts \cite{longhi2019metal,jazaeri2001localization,jiang2019interplay,zeng2019topological,longhi2019topological,ZengQB,SongZ2019}. However, the non-Hermitian effect on the mobility edges in quasicrystals is still lacking. Since the eigenvalues of a non-Hermitian system are generally complex,  particularly interesting questions arise here: whether there exist mobility edges in the non-Hermitian quasiperiodic lattices with short-range or long-range hopping? If so, how we characterize the non-Hermitian mobility edge?

\begin{figure*}
\includegraphics[width=0.88\textwidth]{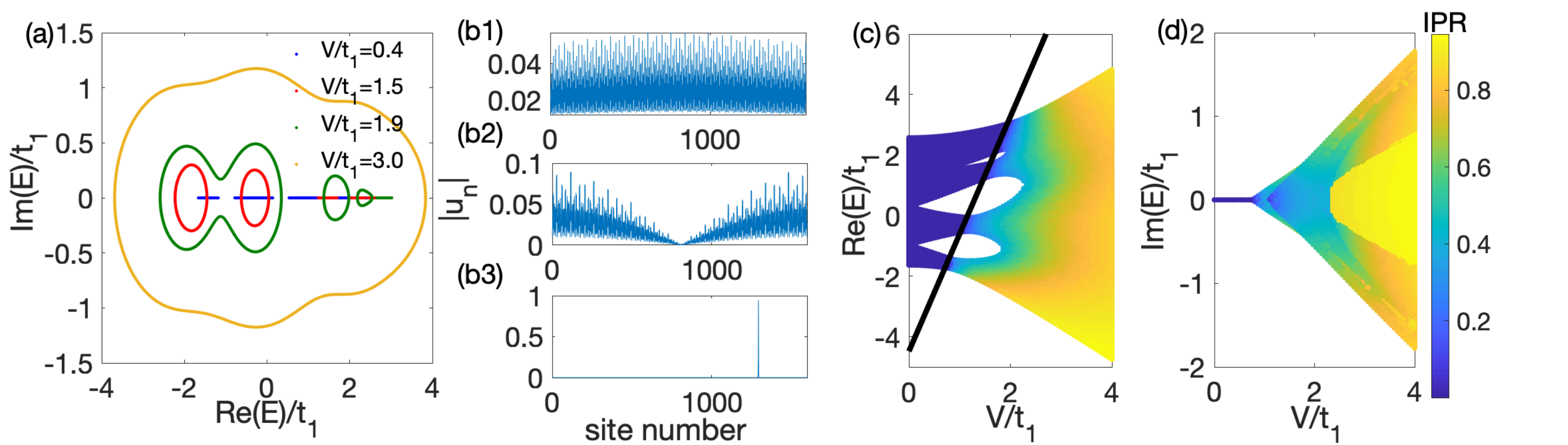}
\caption{Energy eigenvalues and eigenstates of Eq.\ref{Equation01} with lattice sites $L=1597$, $ \alpha=(\sqrt{5}-1)/2 $,
$ p=1.5$ and $ h=0.5$ under PBC. (a) The complex eigenenergies for systems with $V/t_1=0.4,1.5,1.9$, and $3.0$. Distributions of eigenstates
corresponding to different eigenvalues for the system with $ V/t_{1}=1.9 $: in (b1) $  {\rm Re}(E)>{\rm Re}(E_{c}) $ and the corresponding state is an extended state above the mobility
edge, in (b2) $ {\rm Re}(E)\approx {\rm Re}(E_{c}) $ and the state is a critical state near the mobility edge, in (b3) $ {\rm Re}(E)<{\rm Re}(E_{c}) $
and the state is a localized state below the mobility edge. (c) The shading of real energy curves indicates the magnitude of the IPR for
the corresponding eigenstates, and the black solid line represents the boundary given by Eq.(\ref{Equation03}), which separates
localized and extended states. (d) The corresponding imaginary energies of (c). }
\label{fig1}
\end{figure*}

In this work, we address these questions by studying  a non-Hermitian extension of AA model with exponentially decaying short-range hopping and $ \mathcal{PT} $-symmetry.
By analyzing the spatial distribution of wavefunctions and spectral information, we find that the increase of quasiperiodic potential strength can lead a localization transition at the $ \mathcal{PT} $-symmetry breaking point, and unveil that there exists an intermediate regime with mobility edges, which separate the extended states from localized states and are only relevant to real part of spectrum. We also analyze the level statistics and study the Loschmidt echo dynamics of the system.

\section{Generalized non-Hermitian AA model}
We consider a 1D tight binding model with short range hopping terms and a non-Hermitian quasiperiodic potential, described by
\begin{equation}
Eu_n=\sum_{n'\neq n}te^{-p|n-n'|}u_{n'}+V_{n}u_n, \label{Equation01}
\end{equation}
where $t$ is the hopping amplitude with the exponentially decaying parameter $ p>0 $ %where $ u_n $ is the amplitude of wave function at the $ n $th lattice,
and the on site potential $V_{n}$ is given by
\begin{equation}
V_{n}=V\cos(2\pi\alpha n+\phi). \label{Equation02}
\end{equation}
Here $V$ is the potential strength,  $\alpha $ is an irrational number and $ \phi=\theta+ih $ describes a complex phase factor.
When $ h=0$, the model reduces to the Hermitian model studied in Ref.\cite{biddle2011localization,biddle2010predicted}, which is an exponential hopping generalization of
the AA model.
The AA model only includes a nearest-neighbor hopping term with the hopping amplitude
\begin{equation}
t_1=t e^{-p},
\end{equation}
and  manifests a localization-delocalization transition for all eigenstates at the self-dual point $ V=2t_{1} $.
For a finite $ p>0$, the generalized AA model has energy dependent mobility edges given by
$ \cosh(p)=\frac{E+t}{V} $
which was determined by a generalized self-dual
transformation \cite{biddle2010predicted,biddle2011localization}. We note that the transition point and mobility edges are independent of the value of phase factor $\theta$ in the Hermitian limit.

Now we consider the non-Hermitian case with $h \neq 0$. Particularly, we shall consider the case with $ \theta=0 $, for which we have $ V_{n}=V_{-n}^{*}$
and the non-Hermitian model fulfills $ \mathcal{PT} $
symmetry. In the following, we shall study the $ \mathcal{PT} $-symmetric generalized AA (GAA) model with
\begin{equation}
V_{n}=V\cos(2\pi\alpha n+ i h),
\end{equation}
and take $\alpha= (\sqrt{5}-1)/2$ in the whole paper. We note that similar physics is found for $\alpha$ taking other values of irrational numbers.
%For this GAA model, though it is not straightforward to analytically calculate the  exact mobility edge without the self-duality, we can still get the approximate condition for the expression of the real part of  energy dependent mobility edge.
Due to the existence of $ \mathcal{PT}$ symmetry, one may expect that all eigenvalues of the GAA model are real if the $\mathcal{PT}$ symmetry is unbroken. In Fig.\ref{fig1} (a), we display all the eigenvalues of the system with $p=1.5$, $h=0.5$ and various $V$ in the complex space of energies. For convenience, here we take $t_1$ as the unit of energy, and the periodic boundary condition (PBC) is considered. It is shown that all the eigenvalues are real when $V/t_1=0.4$. Further increasing the potential strength $V$ and exceeding a certain threshold $V_{c_1}/t_1=0.702$, the eigenvalues with Re(E) below a critical value $E_c$ become complex accompanying with the breakdown of $ \mathcal{PT}$ symmetry, whereas above $E_c$ remain real, as shown in Fig.\ref{fig1} (a) for $V/t_1=1.5$ and $1.9$. When $V$ exceeds the second threshold  $V_{c_2}/t_1=2.02$, all eigenvalues are complex,  as shown in Fig.\ref{fig1} (a) for $V/t_1=3$.

By inspecting the spatial distribution of the eigenstates, we find that all the states with complex eigenvalues are localized states, whereas the states with real eigenvalues are extended states distributing over the whole lattice. This suggests the  localization transition is simultaneously accompanied by the $ \mathcal{PT}$-symmetry-breaking transition. In Fig.\ref{fig1} (b), we display the distributions of wavefuntions with the real part of eigenvalues Re(E) above, close and below the critical value Re($E_c$) for the system with  $V/t_1=1.9$. It is clear that the state with  Re(E) above the critical value is an extended state and the state below the critical value is a localized state. This indicates clearly that there exists
a regime where the localized and extended states coexist and are separated by mobility edges, when $V$ is in the region $V_{c_1}< V < V_{c_2}$.

\begin{figure}[tbp]
\includegraphics[width=0.45\textwidth]{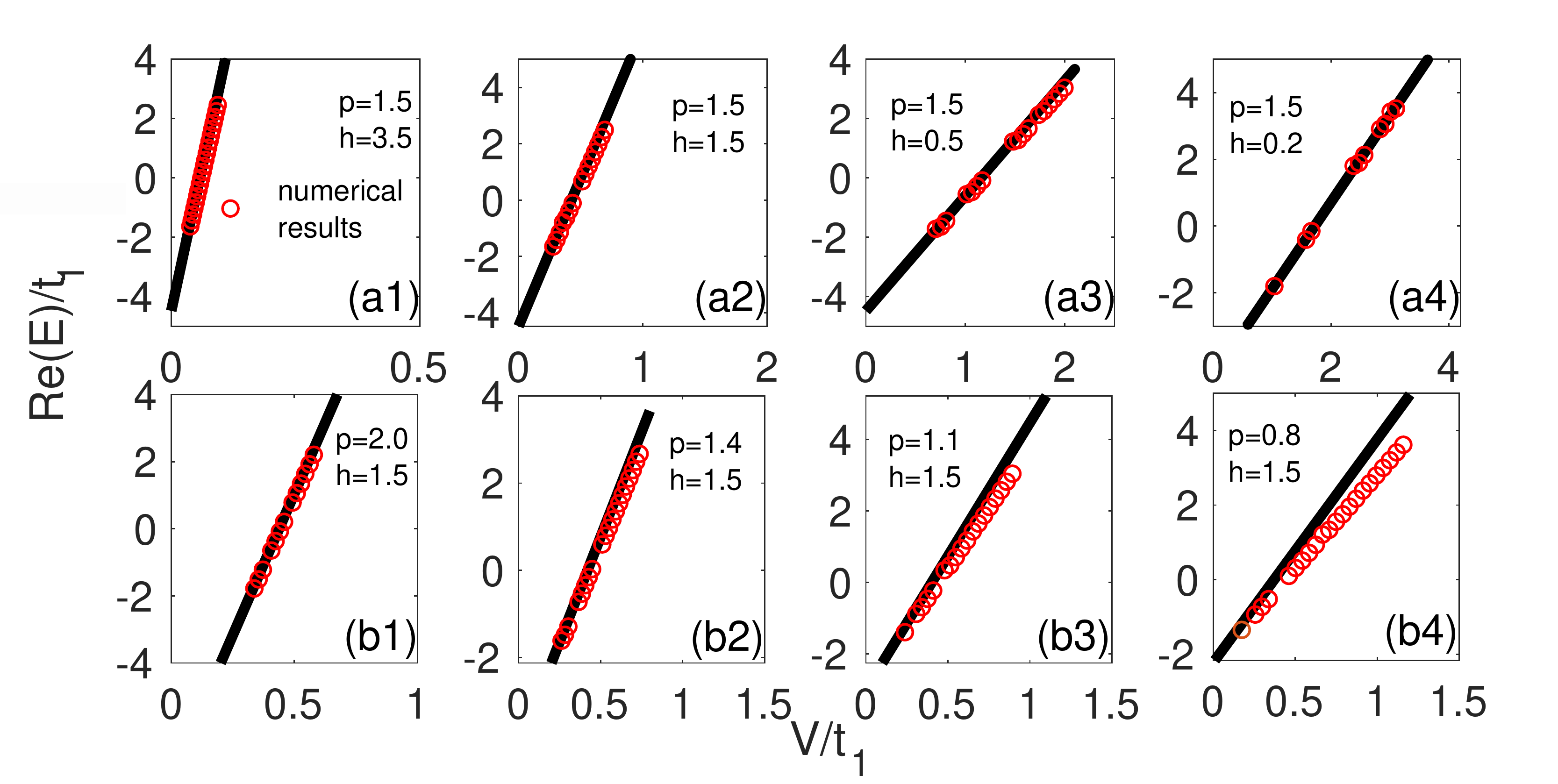}
\caption{Numerical results of mobility edges obtained from IPRs and spectrum (red circles) for systems with $L=1597$ and different parameters (a1)-(a4) $ p=1.5$ and $h=3.5$,  $1.5$,  $0.5 $, and $0.2$, 
(b1)-(b4) $h=1.5 $ and $p=2.0$, $1.4 $, $1.1$ and $0.8$, respectively. The black solid lines are obtained by using Eq.(6).}
\label{fig2}
\end{figure}

Next we determine the mobility edges numerically.
In order to characterize the localization properties of an eigenstate, we calculate the inverse participation ratio (IPR) defined as
\begin{equation}
{\rm IPR}^{(i)}=\frac{\sum_{n}|u_{n}^{i}|^{4}}{(\sum_{n}|u_{n}^{i}|^{2})^{2}}, \label{Equation04}
\end{equation}
where the superscript $ i $ denotes the $ i $th eigenstate of the system, and $n$ labels the lattice coordinate.
Here the corresponding complex
energy $ E_{i} $ is indexed according to their real part $ {\rm Re}(E_{i}) $ in ascending order.
For a full localized
eigenstate, the IPR is finite and IPR$ \simeq 1 $. For an extended state, the IPR $ \simeq1/L $ and approaches
zero when $L$ tends to infinity. In Fig.\ref{fig1}(c) and (d),  we plot the real parts and imaginary parts of eigenvalues as well as the $ \rm{IPR} $ of the corresponding wavefunctions
versus the potential strength $ V$, respectively. The black solid line in the Fig.\ref{fig1}(c) marks the transition points, which separate the extended and localized states, with the values of IPR above which approaching zero and below  being finite. Such a line gives the mobility edge and is found to be well described by a simple relation
\begin{equation}
\cosh(p)=\frac{{\rm Re}(E)+t}{Ve^{h}}. \label{Equation03}
\end{equation} %
Despite lack of exact proof, the above analytical relation for mobility edge boundary agrees well with
numerical results from IPR and spectrum calculations. As shown in Fig.{\ref{fig2}} (a1)-(a4), the numerical results of mobility edges for systems with $p=1.5$ and various $h$ are well described by Eq.(\ref{Equation03}). In Fig.{\ref{fig2}} (b1)-(b4),  we display the numerical results for systems with fixed $h=1.5$ and various $p$. It is shown that  Eq.(\ref{Equation03}) agrees with numerical results for systems with $p=2.5$ and $1.4$, and deviation can be observed for $p=1.1$. From our numerical results, we find that Eq.(\ref{Equation03}) fails to describe SPMEs of systems with $p < 1$ (see  Fig.{\ref{fig2}} (b4) ). When $p$ is small, the effect of long-range hopping becomes more important. Although these systems still support mobility edges, we are not able to get a simple analytical expression for them.

\begin{figure}[tbp]
\includegraphics[width=0.45\textwidth]{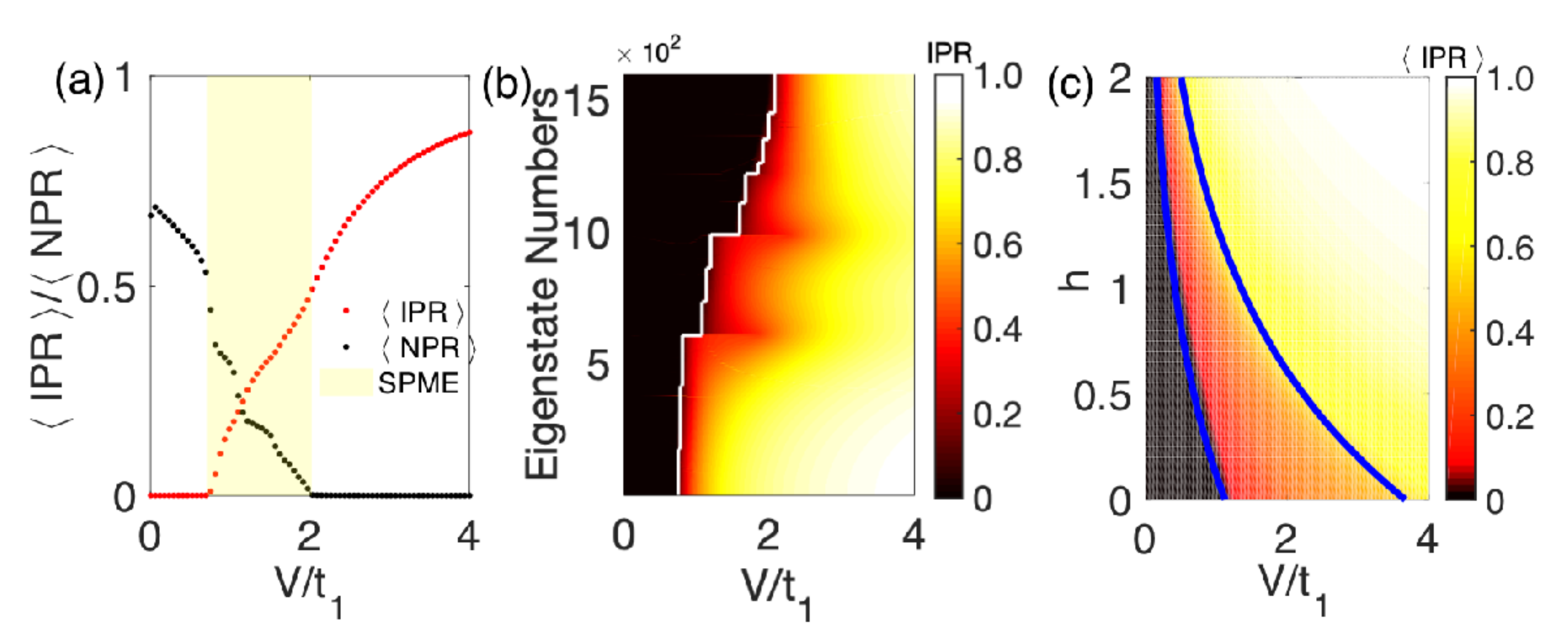}
\caption{Mobility edges for the non-Hermitian GAA model with lattice sites $L=1597$, $ \alpha=(\sqrt{5}-1)/2 $,
and $ p=1.5$. (a) Averaged IPR and NPR for
all eigenstates in our lattice model with $ h=0.5$. (b) IPR of all eigenstates for the system with $ h=0.5$. Here
eigenstates numbers are ordered by $ {\rm Re}(E) $.
The white lines mark out the SPME. (c) Phase diagram in the parameter space spanned by $V/t_1$ and $h$.
The blue solid lines are the phase boundaries separated the intermediate regime from the extended and localized regimes, which can be obtained numerically by using Eq.(\ref{Equation03}).}
\label{fig3}
\end{figure}

%\section{Determination of SPME in non-Hermitian system}
We have become aware of the existence of mobility edges in the non-Hermitian GAA model. To distinguish the region with SPMEs from the extended and localized regions, it is convenient to consider the normalized participation ratio (NPR) defined as
\cite{li2016quantum,li2018mobility,li2017mobility},
\begin{equation}
{\rm NPR}^{(i)}=\left[L\sum_{n}|u_{n}^{i}|^{4}\right]^{-1}, \label{Equation05}
\end{equation}
which is a complementary quantity for the IPR.
%On the contrary for the IPR, the NPR is sensitive to the extended states, and remains finite for spatially delocalized states but vanishes for localized states.
%However, we note that for a single eigenstate the IPR and NPR are trivially interconnected, but
Taking average over all eigenstates, we can get the averaged NPR ($ \left\langle \rm NPR \right\rangle $) and IPR
($ \left\langle \rm IPR \right\rangle $), which provide complete complementary information for the extended, intermediate, and localized
phases. We calculate the NPR and IPR for all eigenstates of the non-Hermitian GAA model and display their average values in
Fig.\ref{fig3}(a),  which shows clearly the existence of three distinct phases depending on the strength of the quasiperiodic potential
$ V/t_{1} $ for the given parameters $ p=1.5 $ and $ h=0.5 $. When the potential strength is smaller than the threshold $ V_{c_1}/t_{1} = 0.702$, all eigenstates are
extended, as indicated by a vanishing $ \left\langle \rm IPR \right\rangle $ and a finite  $ \left\langle \rm NPR \right\rangle $.  When the potential strength exceeds the second threshold $ V_{c_2}/t_{1} = 2.02 $, all eigenstates are localized, as indicated by a finite $ \left\langle \rm IPR \right\rangle $ and a
vanishing $ \left\langle \rm IPR \right\rangle $. When the potential strength lies in between two thresholds, an intermediate regime with
both finite $ \left\langle \rm IPR \right\rangle $ and $ \left\langle \rm NPR \right\rangle $ is characterized by the coexistence of extended and localized states, which can also be read out from the distribution of IPRs for all eigenstates as shown in Fig.\ref{fig3}(b).

We display the average IPR in the two-dimensional parameter space $V/t_1$ versus $h$ in Fig.\ref{fig3}(c), in which the blue solid lines distinguish the extended, intermediate and localized regime, respectively. When gradually increasing $ h $, the intermediate regime with SPME diminishes. On the other hand, if we fix $h$ and increase $ p $, the intermediate regime with SPME also diminishes. Particularly, when $ p \to \infty$, our model reduces to the non-Hermitian AA model with only nearest-neighboring hopping \cite{longhi2019topological,jiang2019interplay}, and  Eq.(\ref{Equation03}) reduces to
\begin{equation}
V e^{h} = 2 t_1,
\end{equation}
indicating the absence of mobility edge.
%In addition to the equilibrium signal for SPME, mobility edge in the non-Hermitian GAA model can result in peculiar dynamical behaves.

\section{Level statistics and Loschmidt echo dynamics}

\begin{figure}[tbp]
\includegraphics[width=0.45\textwidth]{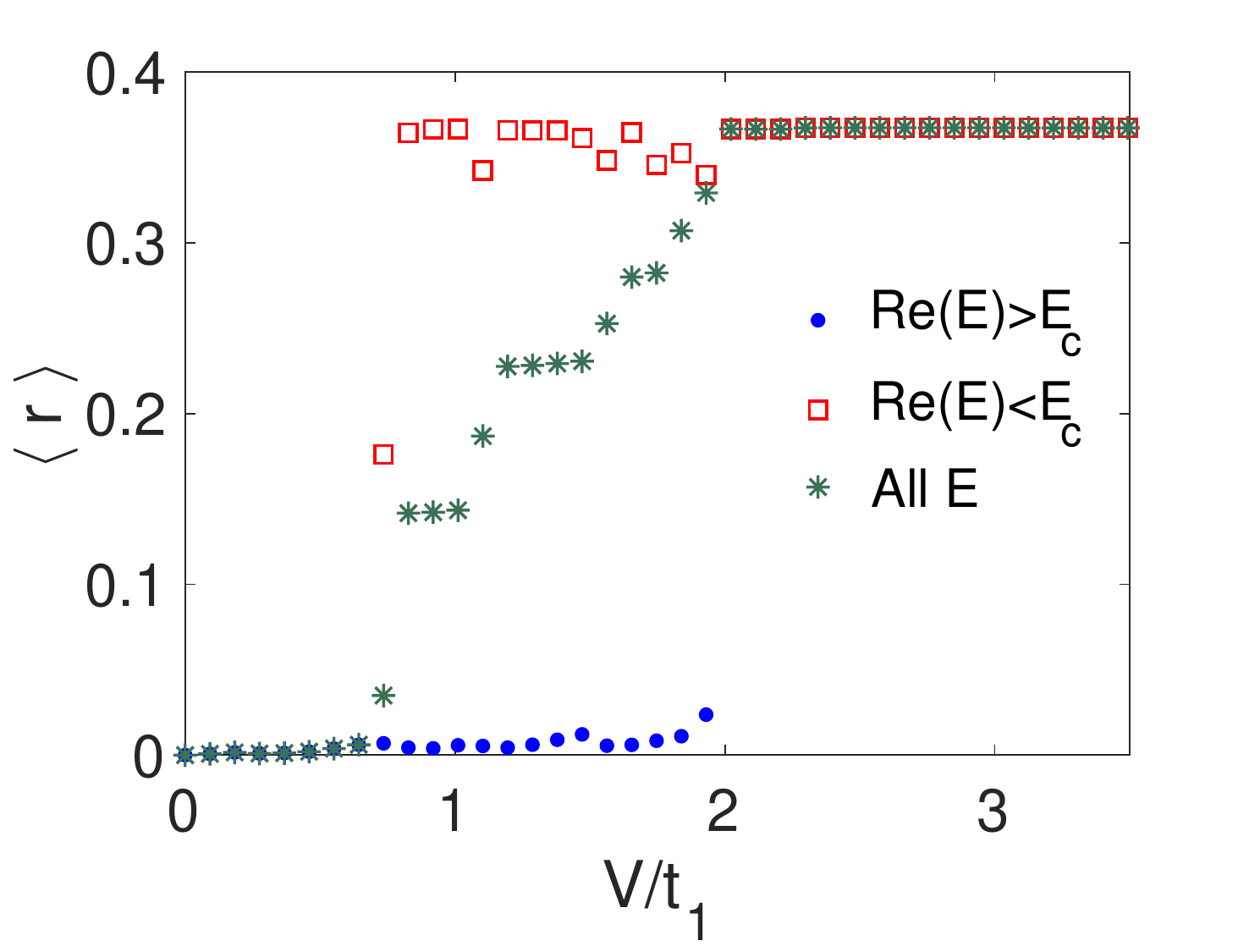}
\caption{The average of adjacent gap ratio $\left\langle r \right\rangle$ for systems with $L=1597$, $ \alpha=(\sqrt{5}-1)/2 $,
$ p=1.5$, $ h=0.5$ and different $V$ versus $V/t_1$.}
\label{fig4}
\end{figure}

The level statistics provides a powerful tool to characterize the localization transition in Hermitian disorder systems \cite{Evangelou,Shore,li2016quantum,Guhr,Huse,WangYC1,WangYC2}.  For our
non-Hermitian model, the eigenvalues in the localized regime are complex. The nearest-neighboring level spacing statistics for non-Hermitian disorder systems has been investigated in terms of non-Hermitian random-matrix theory \cite{Goldsheid,Markum,Molinari,Chalker}. According to Eq.(\ref{Equation03}), the mobility edge is only associated with the real part of complex energies, and it is reasonable to count the real part of level spacings. So, we calculate the adjacent gap ratio $ r $ of ordering $ \Re(E) $, and is given as
\begin{equation}
r_{n}=\frac{{\rm min}(s_{n},s_{n-1})}{{\rm max}(s_{n},s_{n-1})}, \label{Equation06}
\end{equation}
with $ s_{n}$ the level spacing between the real part of the $ n $th and $ (n-1) $th eigenenergies. The average of $ r_{n} $ is introduced as
\begin{equation}
\left\langle r \right\rangle=\frac{1}{L}\sum_{n}r_{n}. \label{Equation07}
\end{equation}
In Fig.\ref{fig4}, we show the real level statistics across the localization transition. The average value $ \left\langle r \right\rangle $
 approaches to zero in the delocalized phase, whereas approaches $ 0.386 $  in the localized phase, which is identical to expected value from Poisson statistics as in the  Hermitian disorder systems.
 For the intermediate regime with mobility edges, the value $ \left\langle r \right\rangle $ presents a
 steplike growth from zero to $ 0.386 $. This is consistent with the result shown in Fig.\ref{fig3}(a). In the intermediate regime, if we count the level statistics for the states above or below the mobility edges separately, the average value $ \left\langle r \right\rangle $
 approaches to the value in the extended or localized regime, respectively, as shown in  Fig.\ref{fig4}.
\begin{figure}[tbp]
\includegraphics[width=0.5\textwidth]{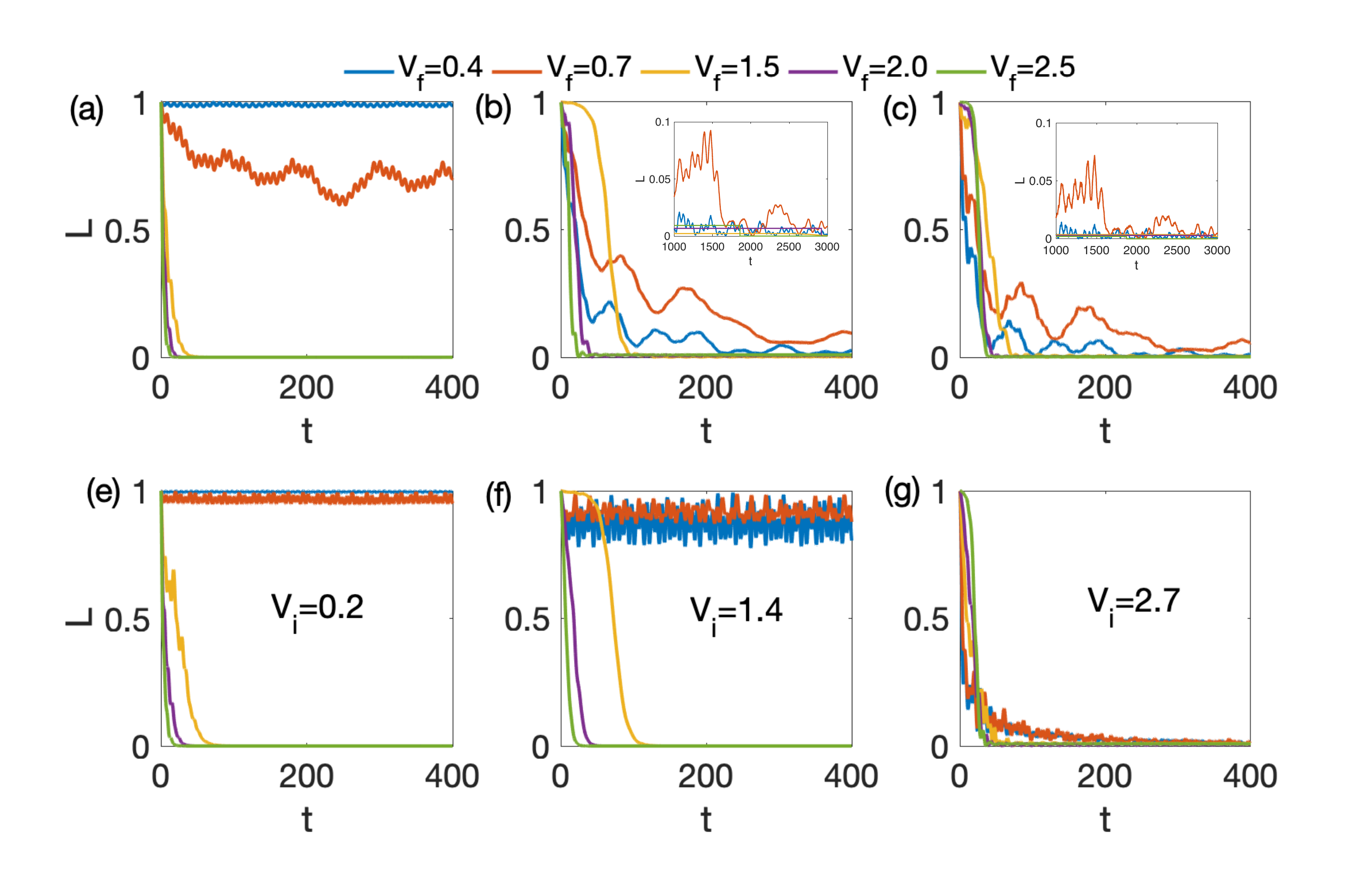}
\caption{Evolution of Loschmidt echo. The initial state is chosen to be the
state corresponding to minimum and maximum of real part of eigenvalues of
the initial Hamiltonian with (a) and (e) $V_{i}=0.2$, (b) and (f) $V_{i}=1.4$, (c) and (g) $V_{i}=2.7$, respectively.
Different $V_{f}$ are shown by different colors. Here we have set the energy unit $t_1=1$.}	
\label{fig5}
\end{figure}

Loschmidt echo is an important quantity for describing quench dynamics \cite{AA16LTP,QuanHT,Heyl,Heyl18RPP,ZhouLW}, which measures the overlap of an initial quantum state and its time-evolution
state after a quench process. The behavior of Loschmidt echo is related to
both the initial state and post-quench states. It was shown that the Loschmidt echo dynamics can
characterize the localization-delocalization transition in standard AA
model \cite{YangC}, and was applied to study the dynamical observation of mobility edges in 1D
incommensurate optical lattices \cite{XuZH}. Here, we explore the Loschmidt-echo
characteristic of our non-Hermitian quasiperiodic system.  The system is initially prepared in the eigenstate
$\left\vert\phi _{i}\right\rangle$ of an initial Hamiltonian $H_i$ with tunable parameter $V= V_i$.
Then the potential strength is suddenly switched to a new value $V_f$, resulting in a final state
\begin{equation}
\left\vert \phi _{f}\left( t\right) \right\rangle =e^{-itH_{f}}\left\vert
\phi _{i}\right\rangle,
\end{equation}
where  $e^{-itH_{f}}$ is the evolution operator after quenching and $\hbar=1$ is set
for convenience. We need to emphasize that for the final system with real eigenvalues,
the final state oscillates over time, and for the final system with complex eigenvalues,
the final state becomes a steady state for a long time, which is similar to the imaginary
time evolution for finding the ground state of a Hermitian system. The difference is that for
the non-Hermitian system the steady state is an eigenstate of the final system with the maximum
 eigenvalue of imaginary part, instead of the ground state.
  The form of Loschmidt echo is
\begin{equation}
L\left( t\right) =\frac{\left\vert \left\langle \phi _{i}|\phi
_{f}\left( t\right) \right\rangle \right\vert ^{2}}{\left\langle \phi
_{i}|\phi _{i}\right\rangle \left\langle \phi _{f}\left( t\right) |\phi
_{f}\left( t\right) \right\rangle },
\end{equation}
where the denominator is introduced to make sure that the initial and final
state are normalized. The dynamics of non-Hermitian system is a kind of 
non-unitary dynamics, due to the existence of complex eigenvalues.

Fig.\ref{fig5}(a) and \ref{fig5}(e) show the quench dynamics for initial states prepared
as eigenstates of the system in the extended regime with $V_i=0.2$, corresponding to minimum and maximum eigenvalues, respectively.
For the final systems with $V_f=0.4$ and $V_f=0.7$, they
locate in the same regime as the initial system with all eigenvalues being real, and $L(t)$ oscillates with a positive lower bound, which
never approaches zero during the evolution process. When the final system locates in the mixed regime with
$V_f=1.5$ and $V_f=2.0$, respectively, both the real and complex eigenvalues coexist,  and $L(t)$ oscillates at short time but approaches zero at
long times. When the final system is in the localized
regime with $V_f=2.7$, $L(t)$ exhibits similar behaviour as in the mixing regime.

Fig.\ref{fig5}(b) and \ref{fig5}(f) show the quench dynamics for initial states prepared
in the mixing regime with $V_i=1.4$, corresponding to minimum and maximum of real part of eigenvalues, respectively.
As one of the initial states is a localized state  and another is an extended state, they exhibit different dynamical behaviors.
While the latter one is similar to the case shown Fig.\ref{fig5}(e), the former one is similar to cases with initial state prepared in the localized regime with $V_i=2.7$ as shown in  Fig.\ref{fig5}(c) and \ref{fig5}(g), where  $L(t)$ always approaches zero at
long times for the finial systems  in different regimes. Our results demonstrate that Loschimt echo exhibits different dynamical behaviors for systems with initial states in different regimes.

\section{Summary}
In summary, we studied localization transition induced by non-Hermitian quasiperiodic potential in 1D $ \mathcal{PT} $-symmetric quasicrystals, described by the non-Hermitian
GAA model with exponential hopping. Our results demonstrate that there exist three different regimes, i.e., extended,  mixed and localized phases. While all the eigenstates are either extended or localized in the extended or localized regime, the extended and localized states coexist in the mixed regime and are separated by  energy dependent mobility edges. By analyzing the distribution of wavefunctions and corresponding eigenenergies, we found that the localization transition is always accompanied by the $ \mathcal{PT} $-symmetry breaking transition and the mobility edges only depend on the real part of energies. We also investigated the level statistics and Loschmidt echo dynamics in our non-Hermitian
quasiperiodic systems and unveiled that they display different behaviors in different regimes.

\begin{acknowledgments}
The work is supported by NSFC under Grants No.11974413 and the National Key
Research and Development Program of China (2016YFA0300600 and
2016YFA0302104).
\end{acknowledgments}

\end{document}